\documentclass[11pt]{iopart}
\usepackage{upgreek}

\usepackage[english]{babel}										% English language/hyphenation
\usepackage[protrusion=true,expansion=true]{microtype}				% Better typography
				
\usepackage[pdftex]{graphicx}									% Enable pdflatex
\usepackage[svgnames]{xcolor}									% Enabling colors by their 'svgnames'
\usepackage[hang, small,labelfont=bf,up,textfont=it,up]{caption}	% Custom captions under/above floats
\usepackage{epstopdf}												% Converts .eps to .pdf
\usepackage{subfig}													% Subfigures
\usepackage{booktabs}												% Nicer tables
\usepackage{fix-cm}

\usepackage{txfonts}

\usepackage{amssymb}
\usepackage{txfonts}
\usepackage{color}
\usepackage{hyperref}

\usepackage{soul}

\usepackage{graphicx}

\usepackage{amssymb}

\usepackage{txfonts}
\usepackage{color}
\usepackage{hyperref}
\hypersetup{
  colorlinks = true,
  citecolor = blue,
  urlcolor = blue}

\usepackage[mathscr]{euscript}

\newcommand{{\bfg}}{\mbox{\boldmath$g$\unboldmath}}
\newcommand{{\bfa}}{\mbox{\boldmath$a$\unboldmath}}
\newcommand{{\bfr}}{\mbox{\boldmath$r$\unboldmath}}
\newcommand{{\bfp}}{\mbox{\boldmath$p$\unboldmath}}
\newcommand{{\bfv}}{\mbox{\boldmath$v$\unboldmath}}
\newcommand{{\bff}}{\mbox{\boldmath$f$\unboldmath}}
\newcommand{{\bfF}}{\mbox{\boldmath$F$\unboldmath}}
\newcommand{{\bfA}}{\mbox{\boldmath$A$\unboldmath}}
\newcommand{{\bfchi}}{\mbox{\boldmath$\chi$\unboldmath}}
\newcommand{{\bfmu}}{\mbox{\boldmath$\mu$\unboldmath}}
\newcommand{{\bfnu}}{\mbox{\boldmath$\nu$\unboldmath}}

\newcommand{{\cA}}{\mbox{\boldmath${\cal A}$\unboldmath}}
\newcommand{{\cJ}}{\mbox{\boldmath${\cal J}$\unboldmath}}
\newcommand{{\cF}}{\mbox{\boldmath${\cal F}$\unboldmath}}
\newcommand{{\cG}}{\mbox{\boldmath${\cal G}$\unboldmath}}
\newcommand{{\cE}}{\mbox{\boldmath${\cal E}$\unboldmath}}
\newcommand{{\cB}}{\mbox{\boldmath${\cal B}$\unboldmath}}
\newcommand{{\cX}}{\mbox{\boldmath${\cal X}$\unboldmath}}
\newcommand{{\cY}}{\mbox{\boldmath${\cal Y}$\unboldmath}}

\def\v#1{{\bf#1}}

\newcommand{{\mA}}{\mbox{\boldmath${\mathscr{A}}$\unboldmath}}
\newcommand{{\mZ}}{\mbox{\boldmath${\mathscr{Z}}$\unboldmath}}
\newcommand{{\mF}}{\mbox{\boldmath${\mathscr{F}}$\unboldmath}}
\newcommand{{\mG}}{\mbox{\boldmath${\mathscr{G}}$\unboldmath}}
\newcommand{{\mE}}{\mbox{\boldmath${\mathscr{E}}$\unboldmath}}
\newcommand{{\mB}}{\mbox{\boldmath${\mathscr{B}}$\unboldmath}}
\newcommand{{\mX}}{\mbox{\boldmath${\mathscr{X}}$\unboldmath}}
\newcommand{{\mY}}{\mbox{\boldmath${\cal Y}$\unboldmath}}

\usepackage[top= 2.4 cm, bottom= 2.6 cm, left=2.9 cm, right=2.9 cm]{geometry}

\begin{document}

\title[{\rm J A Heras and R Heras}]{{\fontfamily{qag}\selectfont {\LARGE \textcolor[rgb]{0.00,0.00,0.49}{Helmholtz's theorem for two retarded fields and its application to Maxwell's equations}}}}

\vskip 30pt

\author{Jos\'e A. Heras$^1$ and Ricardo Heras$^2$}
\address{$^1$Instituto de Geof\'isica, Universidad Nacional Aut\'onoma de M\'exico, Ciudad de M\'exico 04510, M\'exico.
E-mail: herasgomez@gmail.com\\
$^2$Department of Physics and Astronomy, University College London, London WC1E 6BT, UK. E-mail: ricardo.heras.13@ucl.ac.uk }
\begin{abstract}
\noindent An extension of the Helmholtz theorem is proved, which states that two retarded vector fields ${\bf F}_1$ and ${\bf F}_2$
satisfying appropriate initial and boundary conditions are uniquely determined by specifying their divergences $\nabla\cdot{\bf F}_{1}$ and $\nabla\cdot{\bf F}_{2}$ and their coupled curls $-\nabla\times{\bf F}_{1}-\partial {\bf F}_{2}/\partial t$ and $\nabla\times{\bf F}_{2}-(1/c^2)\partial {\bf F}_{1}/\partial t$, where $c$ is the propagation speed of the fields. When a corollary of this theorem is applied to Maxwell's equations, the retarded electric and magnetic fields are directly obtained. The proof of the theorem relies on a novel demonstration of the uniqueness of the solutions of the vector wave equation.
\end{abstract}
\vskip 30pt

\section*{{{\fontfamily{qag}\selectfont {\large \textcolor[rgb]{0.00,0.00,0.49}{1. Introduction}}}}}

 The mathematical foundations of electrostatics and magnetostatics relies on the Helmholtz theorem of vector analysis.
Among the several equivalent formulations of this theorem presented in standard textbooks \cite{1,2,3,4}, let us consider the formulation given by Griffiths \cite{1}. The theorem states that if the divergence $\nabla\cdot\v F = D(\v r)$ and the curl $\nabla\times\v F=\v C(\v r)$ of a vector function $\v F(\v r)$ are specified, and if they both go to zero faster than $1/r^2$ as $r\rightarrow\infty,$ and if $\v F(\v r)$ goes to zero as $r\rightarrow\infty,$ then  $\v F$ is uniquely given by $\v F = - \nabla U  + \nabla\times \v W,$ where
\begin{eqnarray}
 U(\v r)\equiv \frac{1}{4\pi}\int \frac{D ( \v r') }{R} \, d^3r',\;\; W(\v r)\equiv \frac{1}{4\pi}\int \frac{\v C ( \v r') }{R}\, d^3r'.
\end{eqnarray}
Here $\v r$ is the field point and $r=|\v r|$ is its magnitude, $\v r'$ is the source point and $R=\!|\v r-\v r'|$. The integrals are over all space and
 $d^3r'$ is the volume element. This theorem has a useful corollary: Any vector function $\v F(\v r)$ that goes to zero faster than $1/r$ as $r\rightarrow\infty$ can be expressed as
\begin{eqnarray}
\v F (\v r)= -\nabla\bigg(\frac{1}{4\pi}\int \frac{\nabla' \cdot \v F ( \v r') }{R} \, d^3r'\bigg)  + \nabla\times \bigg(\frac{1}{4\pi}
\int \!\frac{\nabla' \times \v F ( \v r') }{R}\, d^3r'\bigg).
\end{eqnarray}
In a first application, we write $\v F=\v E$ and use the electrostatic equations: $\nabla \cdot \v E=\rho/\epsilon_0$ and $\nabla \times \v E=0$ to obtain the electrostatic field $\v E(\v r)$ produced by the charge density $\rho(\v r)$:
\begin{eqnarray}
\v E (\v r)= -\nabla\bigg(\frac{1}{4\pi\epsilon_0}\int \frac{\rho(\v r') }{R}\, d^3r'\bigg),
\end{eqnarray}
or equivalently $\v E=-\nabla \Phi,$ where $\Phi$ is the scalar potential:
\begin{eqnarray}
\Phi(\v r)= \frac{1}{4\pi\epsilon_0}\int \frac{\rho(\v r') }{R}\, d^3r'.
\end{eqnarray}
 In a second application we make $\v F=\v B$ and use the magnetostatic  equations: $\nabla\cdot \v B=0$ and $\nabla \times \v B=\mu_0\v J$ to obtain the magnetostatic field $\v B(\v r)$ produced by the current density $\v J(\v r)$:
\begin{eqnarray}
\v B (\v r)= \nabla \times\bigg(\frac{\mu_0}{4\pi}\int \frac{\v J(\v r')}{R}\, d^3r'\bigg),
\end{eqnarray}
or equivalently $\v B=\nabla\times \v A,$ where $\v A$ is the vector potential:
\begin{eqnarray}
\v A (\v r)= \frac{\mu_0}{4\pi}\int \frac{\v J(\v r')}{R}\, d^3r'.
\end{eqnarray}
Here $\epsilon_0$ and $\mu_0$ are the permittivity and permeability of vacuum which satisfy $\epsilon_0\mu_0=1/c^2$, with $c$ being the speed of light in vacuum.

The Helmholtz theorem is also applicable to the time-dependent regime of Maxwell's equations \cite{5,6}. The reason is simple: the derivation of (2) does not involve time and therefore it can be applied to a time-dependent vector field $\v F(\v r,t)$. We simply make the replacements $\v F(\v r)\!\to\! \v F(\v r,t)$ and $\v F(\v r')\!\to\! \v F(\v r',t)$ in (2) and obtain an {\it instantaneous} form of the theorem \cite{6}:
\begin{eqnarray}
\v F (\v r, t)\!= -\nabla\bigg(\frac{1}{4\pi}\int \frac{\nabla' \cdot \v F ( \v r',t) }{R} \,d^3r'\bigg)  + \nabla\times \bigg(\frac{1}{4\pi}
\int \!\frac{\nabla' \times \v F ( \v r',t) }{R}\, d^3r'\bigg).
\end{eqnarray}
 However, this form of the theorem is of {\it limited} practical utility in the time-dependent regime of Maxwell's equations because  the Faraday law
and the Amp$\grave{\rm e}$re-Maxwell law in this regime do not specify the curls of the fields $\v E$ and $\v B$ as such in terms of sources, rather they specify the {\it hybrid} quantities $\nabla\times \v E+\partial \v B/\partial t$ and $\nabla\times \v B-(1/c^2)\partial \v E/\partial t$, which couple space and time variations of both fields. In the more general case in which there are magnetic monopoles, the Faraday
and Amp$\grave{\rm e}$re-Maxwell laws read $-\nabla\times \v E-\partial \v B/\partial t=\mu_0\v J_g$ and $\nabla\times \v B-(1/c^2)\partial \v E/\partial t=\mu_0 \textbf{J}_e$, where $\textbf{J}_g$ is  the magnetic current density and $\textbf{J}_e$ is  the electric current density.  As may be seen, these laws connect the hybrid quantities $-\nabla\times \v E-\partial \v B/\partial t$ and $\nabla\times \v B-(1/c^2)\partial \v E/\partial t$ with their respective electric and magnetic currents. It is clear that the time derivatives of the fields couple their corresponding curls. Let us call the first hybrid quantity the \emph{coupled curl} of $\v E$ and the second quantity the \emph{coupled curl} of $\v B$.

It would be desirable to have an extension of the Helmholtz theorem which allows us to directly find the retarded electric and magnetic fields in terms of the retarded scalar and vector  potentials, in the same form that the standard Helmholtz theorem allows us to directly find the electrostatic and magnetostatic fields in terms of their respective static scalar and vector potentials. Expectably, this appropriate generalisation of the theorem should be formulated for two retarded vector fields and in terms of their respective divergences and coupled curls.

In this paper we formulate a generalisation of the Helmholtz theorem for two retarded vector functions. We show how a corollary of this theorem allows us to find the retarded electric and magnetic fields and introduce the corresponding retarded potentials in three specific cases: for Maxwell equations with the electric charge and current densities (the standard case), when these equations additionally have polarisation and magnetisation densities and finally when they additionally have magnetic charge and current densities. The proof of the theorem relies on the uniqueness of the solutions of the vector wave equation. A novel demonstration of this uniqueness is presented in the Appendix A. Although more complicated than the standard Helmholtz theorem, the extension of this theorem for two retarded vector fields formulated here may be presented in an advanced undergraduate course of electrodynamics.

\section*{{{\fontfamily{qag}\selectfont {\large \textcolor[rgb]{0.00,0.00,0.49}{2. The Helmholtz theorem for two retarded fields}}}}}

\noindent We begin by defining a {\it retarded} function as one function of space and time whose sources are evaluated at the retarded time.
For example, the vector $\cF(\v r,t)= \int \bfg( \v r',t-R/c )d^3r'$
is a retarded field because its vector source $\bfg$ is evaluated at the retarded time $t-R/c$. In order to simplify the notation, we will use the retardation brackets $[\quad]$ to indicate that the enclosed quantity is to be evaluated at the source point $\v r'$ and at the retarded time $t-R/c,$ that is, $[\cF]=\cF(\v r',t-R/c)$. For example, $\cF= \int [\bfg]d^3r'$ denotes a retarded quantity.

Suppose we are told that the divergences $\nabla\cdot\v F_1$ and $\nabla\cdot\v F_2$ of two retarded vector functions $\v F_1=\v F_1(\v r,t)$ and $\v F_2=\v F_2(\v r,t)$ are  specified by the time-dependent scalar functions $D_1=D_1(\v r,t)$ and  $D_2=D_2(\v r,t)$:
\begin{eqnarray}
\nabla\cdot\v F_1=D_1, \qquad \nabla\cdot\v F_2=D_2,
\end{eqnarray}
and that the coupled curls  $-\nabla\times\v F_1-\partial \v F_2/\partial t$ and $\nabla\times\v F_2-(1/c^2)\partial \v F_1/\partial t$ are specified by the time-dependent vector functions $\v C_1=\v C_1(\v r,t)$ and $\v C_2=\v C_2(\v r,t)$:
\begin{eqnarray}
-\nabla\times\v F_1-\frac{\partial \v F_2}{\partial t}=\v C_2, \qquad \nabla\times\v F_2-\frac{1}{c^2}\frac{\partial \v F_1}{\partial t}=\v C_1,
\end{eqnarray}
where $c$ is the propagation speed of the fields $\v F_1$ and $\v F_2$.\footnote[1]{Here the speed $c$ does not necessarily have to represent the speed of light. However, if $c$ is identified with the speed of light in vacuum and the following identifications: $\v F_1=\v E,\v F_2=\v B, D_1=\rho/\epsilon_0,  D_2=0, \v C_1=\mu_0\v J$ and $\v C_2=0$ with $\epsilon_0\mu_0=1/c^2$ are made, then (8) and (9) become Maxwell's equations in SI units. Analogously, if the identification $\v F_1=c\v E,\v F_2=\v B, D_1=4\pi c \rho,  D_2=0, \v C_1=(4\pi/c)\v J$ and $\v C_2=0$ are made then (9) and (10) become Maxwell's equations in Gaussian units.} For consistence, $\v C_1, \v C_2, D_1$ and $D_2$ must satisfy
the ``continuity'' equations:
\begin{eqnarray}
\nabla\cdot\v C_1+\frac{1}{c^2}\frac{\partial D_1}{\partial t}=0, \qquad \nabla\cdot\v C_2+\frac{\partial D_2}{\partial t}=0,
\end{eqnarray}
which are implied by (8) and (9). {\it Question}: Can we, on the basis of this information, uniquely determine $ \v F_1$ and $\v F_2$? If $D_1,D_2,\v C_1$ and $\v C_2$ go to zero sufficiently rapidly at infinity, the answer is $yes$, as we will show by explicit
construction.\footnote[2]{The answer is also $yes$ for the case in which the quantities $D_1,D_2,\v C_1$ and $\v C_2$ are localised in space, that is, when they are zero outside a finite region of space. This frequently occurs  in practical applications because these quantities usually play the role of sources, which are physically localised in space.
}

We claim that the solution of (8) and (9) is
\begin{eqnarray}
\v F_1 = - \nabla U_1  - \nabla\times \v W_1 - \frac{\partial \v W_2}{\partial t},\;\;\;  \v F_2 = - \nabla U_2  + \nabla\times \v W_2 - \frac{1}{c^2}\frac{\partial \v W_1}{\partial t} ,
\end{eqnarray}
where
\begin{eqnarray}
U_1= {1\over 4\pi}\int \frac{[D_1]}{R} \,d^3r',\qquad U_2= {1\over 4\pi}\int {[D_2]\over R} ]\,d^3r',
\end{eqnarray}
and
\begin{eqnarray}
\v W_1= {1\over 4\pi}\int {[\v C_2]\over R} \,d^3r',\qquad \v W_2= {1\over 4\pi}\int {[\v C_1]\over R} \,d^3r'.
\end{eqnarray}
The integrals are over all space. Our demonstration requires two {\it additional} sets of equations. The first set is formed by the  ``Lorenz'' conditions:
\begin{eqnarray}
\nabla\cdot\v W_1+\frac{\partial U_2}{\partial t}=0,\qquad \nabla\cdot\v W_2+\frac{1}{c^2}\frac{\partial U_1}{\partial t}=0.
\end{eqnarray}
We will prove the first condition using the result \cite{2}: $\nabla\cdot\big([\cF]/R\big)\!+\!\nabla'\cdot\big([\cF]/R \big)\!=\![\nabla'\cdot\cF]/R$,
\begin{eqnarray}
\nabla\cdot\v W_1&=\frac{1}{4\pi}\int\nabla\cdot\bigg(\frac{[\v C_2]}{R}\,d^3r'\bigg)\nonumber\\
                 &= \frac{1}{4\pi}\int \frac{[\nabla'\cdot\v C_2]}{R}\,d^3r'-\frac{1}{4\pi}\int\nabla'\cdot\bigg(\frac{[\v C_2]}{R}\bigg)\,d^3r'\nonumber\\
                 &=-\frac{\partial }{\partial t}\bigg(\frac{1}{4\pi}\int\frac{[D_2]}{R}\,d^3r'\bigg) -\frac{1}{4\pi}\oint\frac{\hat{\v n}\cdot[\v C_2]}{R}\,dS=-\frac{\partial U_2 }{\partial t},
\end{eqnarray}
where we have used the second equation given in (10), the property \cite{7}:  $[\partial\cF/\partial t]=\partial[\cF]/\partial t$, the Gauss theorem to transform the second volume integral of the second line into a surface integral ($dS$ is the surface element), which is seen to vanish at infinity by assuming the boundary condition that $\v C_2$  goes to zero faster than $1/r^2$ as $r\rightarrow\infty$, and finally considering the second equation in (12).\footnote[3]{In physical applications, the vector $\v C_2$ usually represents a localised quantity. This means that $\v C_2$ is different from zero only within a limited region of space. However, the surface of integration $S$ in the surface integral on the third line of (15) encloses all space, and therefore $S$ is outside the region where $\v C_2$ is different from zero. It follows that $\v C_2$ is zero everywhere on $S$ and then the surface integral vanishes.} Following a similar procedure, we can prove the second ``Lorenz'' condition given in (14). The required second set of equations is formed by the wave equations
\begin{eqnarray}
\Box^2U_1=-D_1,\quad \square^2U_2=-D_2,\quad \square^2\v W_1=-\v C_2,\quad \square^2\v W_2=-\v C_1,
\end{eqnarray}
where $\Box^{2}\equiv\nabla^2-(1/c^2)\partial^2/\partial t^2$ is the d'Alembertian operator. Using the result \cite{7}:
\begin{eqnarray}
\Box^2\bigg(\frac{[{\cX}]}{R}\bigg)=-4\pi[{\cX}]\delta(\v r-\v r'),
\end{eqnarray}
where $\cX$ represents a scalar or vector function,\footnote[4]{This identity is true for functions ${\cX}$ such that the quantities $[{\cX}]/R$ have not the form  $[{\cX}]/R=f(R)[\v F]$ with $f(R)$ being a polynomial function. If for example $f(R)=R$ then
$[{\cX}]/R=R[\v F]$. It follows that $\Box^2 (R[\v F])=-4\pi R[\v F]\delta(\v r-\v r')=0$ since the factor $R\delta(\v r-\v r')$
vanishes for $\v r\not=\v r'$ because of the delta function and also for $\v r=\v r'$  because this equality implies $R=0$.} we can show the first wave equation in (16),
\begin{eqnarray}
\square^2 U_1 =\frac{1}{4\pi}\int\! \square^2\bigg(\frac{[D_1]}{R}\bigg)\,d^3r'=-\int\![D_1]\delta(\v r-\v r')\,d^3r'=-D_1.
\end{eqnarray}
By an entirely similar procedure, we can show the remaining wave equations given  in (16). Considering (11), (14) and (16), we can now prove that $\v F_1$ and $\v F_2$ satisfy (8) and (9). We take the divergences to $\v F_1$ and $\v F_2$ and obtain
\begin{eqnarray}
\nabla\cdot\v F_1&=-\nabla^2U_1-\frac{\partial}{\partial t}\nabla\cdot\v W_2=-\nabla^2U_1+\frac{1}{c^2}\frac{\partial^2 U_1}{\partial t^2}=-\Box^2U_1=D_1,\\
\nabla\cdot\v F_2&=-\nabla^2U_2-\frac{1}{c^2}\frac{\partial}{\partial t}\nabla\cdot\v W_1=-\nabla^2U_1+\frac{1}{c^2}\frac{\partial^2 U_2}{\partial t^2}=-\Box^2U_2=D_2.
\end{eqnarray}
We now calculate the coupled curls of $\v F_1$ and $\v F_2$ and obtain
\begin{eqnarray}
-\nabla\times\v F_1-\frac{\partial \v F_2}{\partial t} =\;\nabla(\nabla\cdot\v W_1)-\nabla^2\v W_1+\frac{\partial}{\partial t}\nabla\times \v W_2
\nonumber\\
\qquad\qquad\qquad+\nabla\bigg(\frac{\partial U_2}{\partial t }\bigg)-\nabla\times \frac{\partial \v W_2}{\partial t }+\frac{1}{c^2}\frac{\partial^2 \v W_1}{\partial t^2}\nonumber \\
\qquad\qquad\quad\;\;=\;\nabla\bigg(\nabla\cdot\v W_1+\frac{\partial U_2}{\partial t}\bigg)-\square^2\v W_1=-\square^2\v W_1=\v C_2,\\
\;\;\nabla\times\v F_2-\frac{\partial \v F_1}{\partial t} =\;\nabla(\nabla\cdot\v W_2)-\nabla^2\v W_2+\frac{\partial}{\partial t}\nabla\times \v W_1
\nonumber\\
\qquad\qquad\qquad+\nabla\bigg(\frac{\partial U_1}{\partial t }\bigg)-\nabla\times \frac{\partial \v W_1}{\partial t }+\frac{1}{c^2}\frac{\partial^2 \v W_2}{\partial t^2}\nonumber \\
\qquad\qquad\quad\;\;=\;\nabla\bigg(\nabla\cdot\v W_2+\frac{\partial U_1}{\partial t}\bigg)-\square^2\v W_2=-\square^2\v W_2=\v C_1.
\end{eqnarray}
Therefore the functions $\v F_1$ and $\v F_2$ given in (11) constitute a solution of (8) and (9).

Now, assuming that the conditions on $D_1,D_2,\v C_1$ and $\v C_2$ are met, is the solution in (11) $unique$? The answer is clearly $no$, for we can add to $\v F_1$ the function $\v H_1$ and $\v F_2$ the function $\v H_2$, with the divergences and coupled curls of $\v H_1$ and $\v H_2$ being zero, and the result still has divergences $D_1$ and $D_2$ and coupled curls $\v C_1$ and $\v C_2$. However, it so happens that if $\v F_1, \v F_2, \partial\v F_1/\partial t$ and
$ \partial\v F_2/\partial t$ vanish at $t=0$, and if  $\nabla \v F_1,\nabla \v F_2, \partial\v F_1/\partial t$ and
$ \partial\v F_2/\partial t$ go to zero faster than $1/r^2$ as $r\rightarrow\infty$, and if $\v F_1$ and $\v F_2$ themselves go to zero as $r\rightarrow\infty$,
then $\v H_1$ and $\v H_1$ are zero and therefore the solution (11) is unique. In the notation we are using a generic  second-order tensor $\nabla\cF$ denotes the gradient of the vector $\cF$. The explicit proof that $\v H_1$ and $\v H_1$ are zero under their specified initial and boundary conditions is presented in Appendix A. Notice that the conditions that $\nabla \v F_1,\nabla \v F_2, \partial\v F_1/\partial t$ and
$ \partial\v F_2/\partial t$ go to zero faster than $1/r^2$ as $r\rightarrow\infty$ {\it automatically} imply that $D_1,D_2,\v C_1$ and $\v C_2$ go to zero faster than $1/r^2$ as $r\rightarrow\infty$.
\vskip 5pt
Now we can state the Helmholtz theorem for two retarded fields more rigorously:
\vskip 5pt
\noindent \textbf{Theorem.} If the divergences $D_1$ and $D_2$ and the coupled curls $\v C_1$ and $\v C_2$ of two retarded vector functions $\v F_1$ and $\v F_2$ are specified, and if $\v F_1,\v F_2, \partial\v F_1/\partial t$ and
$ \partial\v F_2/\partial t$ vanish at $t=0$, and if $\nabla \v F_1,\nabla \v F_2, \partial\v F_1/\partial t$ and
$ \partial\v F_2/\partial t$ go to zero faster than $1/r^2$ as $r\rightarrow\infty$, and if $\v F_1$ and $\v F_2$ themselves go to zero as $r\rightarrow\infty$, then $\v F_1$ and $\v F_2$ are uniquely given by (11).
\vskip 5pt
This theorem has the following corollary:
\noindent The fields $\v F_1$ and $\v F_2$ can be expressed as
\begin{eqnarray}
\v F_1 =& - \nabla \bigg\{\int\!\frac{[\nabla'\cdot\v F_1]}{4\pi R} \,d^3r'\bigg\}
- \nabla \times \bigg\{\int\!\frac{[-\nabla'\!\times\v F_1-\partial \v F_2/\partial t]}{4\pi R} \,d^3r'\bigg\}\nonumber\\
& - \frac{\partial}{\partial t}\bigg\{\int\!\frac{[\nabla'\!\times\v F_2-(1/c^2)\partial \v F_1/\partial t]}{4\pi R} \,d^3r'\bigg\},\\
\v F_2 =& - \nabla \bigg\{\int\!\frac{[\nabla'\cdot\v F_2]}{4\pi R} \,d^3r'\bigg\}
+\nabla \times\! \bigg\{\int\!\frac{[\nabla'\!\times\v F_2-(1/c^2)\partial \v F_1/\partial t]}{4\pi R} \,d^3r'\bigg\}\nonumber\\
& - \frac{\partial}{\partial t}\bigg\{\int\!\frac{[-\nabla'\!\times\v F_1-\partial \v F_2/\partial t]}{4\pi R} \,d^3r'\bigg\}.
 \end{eqnarray}
These expressions of the Helmholtz theorem for two retarded vector functions are considerably more complicated than the expression of the Helmholtz theorem (2) for a static vector function. However, the practical advantage (23) and (24) is that they allows us to directly find the retarded fields of Maxwell's equations, as we will see in the next section.

\section*{{{\fontfamily{qag}\selectfont {\large \textcolor[rgb]{0.00,0.00,0.49}{3. Applications }}}}}
 \noindent In a first application of the corollary expressed in (23) and (24), we make the identifications:
$\v F_1=\v E$ and $\v F_2=\v B$ in (23) and (24) and subsequently use the Maxwell equations in SI units
\begin{eqnarray}
\nabla\cdot\v E=\,\frac{1}{\epsilon_0}\rho,\qquad \qquad\quad\;\;\nabla\cdot\v B= 0, \\
\nabla\times \v E+\frac{\partial \v B}{\partial t}=0,\quad
\nabla\times \v B-\frac{1}{c^2}\frac{\partial \v E}{\partial t}= \mu_0\v J,
\end{eqnarray}
to directly obtain the familiar retarded fields
\begin{eqnarray}
\v E =& - \nabla \bigg({1\over 4\pi\epsilon_0}\int\!\frac{[\rho]}{R} \,d^3r'\bigg)
- \frac{\partial}{\partial t}\bigg (\frac{\mu_0}{4\pi}\int\frac{[\v J]}{R}\,d^3r'\bigg),\\
\v B =& \nabla \times \bigg(\frac{\mu_0}{4\pi}\int\frac{[\v J]}{R}\,d^3r'\bigg),
 \end{eqnarray}
or more compactly,
\begin{eqnarray}
\v E = - \nabla \Phi - \frac{\partial \v A}{\partial t},\quad\v B =\nabla \times \v A,
 \end{eqnarray}
 where we have defined the retarded scalar and vector potentials as
 \begin{eqnarray}
\Phi=& {1\over 4\pi\epsilon_0}\int\!\frac{[\rho]}{R}\,d^3r',\quad \v A=\frac{\mu_0}{4\pi}\int\frac{[\v J]}{R}\,d^3r'.
 \end{eqnarray}
Here $c$ is the speed of light in vacuum and is defined by $c=1/\sqrt{\epsilon_0\mu_0}$.\footnote[5]{It is interesting to note that (23) and (24) can be used to find the retarded fields of the gravitational theory suggested by Heaviside \cite{8}, which is similar in form to that of Maxwell.
If we write
$\v F_1=\v g$ and $\v F_2=\v k$, where the time-dependent vector fields $\v g$ and $\v k$ are respectively the gravitoelectric and gravitomagnetic fields, use equations (23) and (24), and consider the gravitational equations in SI units: $\nabla\cdot\v g=-\rho_{\textsc{g}}/\tilde{\epsilon_0},\nabla\cdot\v k= 0,
\nabla\times \v g+\partial \v k/\partial t=0, \nabla\times \v k-(1/c^2)\partial \v g/\partial t= -\tilde{\mu}_0\v J_{\textsc{g}},$
where $\rho_{\textsc{g}}$ and $\v J_{\textsc{g}}$ are the mass density and the mass current density, which satisfy the continuity equation $\nabla\cdot\v J_{\textsc{g}}+\partial\rho_{\textsc{g}}/\partial t=0$ (which represents the mass conservation), then we get the retarded fields $\v g = -\nabla V_{\textsc{g}} - \partial \v A_{\textsc{g}}/\partial t$ and $\v k =\nabla \times \v A_{\textsc{g}}$,
 where the retarded scalar and vector potentials are
 \[
V_{\textsc{g}}=- {1\over 4\pi\tilde{\epsilon}_0}\int\!\frac{[\rho_{\textsc{g}}]}{R} \,d^3r',\quad \v A_{\textsc{g}}=-\frac{\tilde{\mu}_0}{4\pi}\int\frac{[\v J_{\textsc{g}}]}{R}\,d^3r'.
 \]
Here the gravitational permittivity constant of vacuum is defined as $\tilde{\epsilon}_0=1/(4\pi \textsc{G})$, where $\textsc{G}$ is the universal gravitational constant. The gravitational permeability constant of vacuum is defined as $\tilde{\mu}_0=4\pi\mathfrak{S}/c^2$. It follows that $\tilde{\epsilon}_0\tilde{\mu}_0=1/c^2$, where $c$ is the speed of light in vacuum. Jefimenko \cite{9}, McDonald \cite{10} and recently,  Vieira and Brentan \cite{11} have discussed this gravitational theory, whose equations have also been obtained using an alternative approach \cite{12}.}

 In a second application of (23) and (24) , we make the identifications:
$\v F_\texttt{1}=\v E$ and $ \v F_{\texttt{2}}=\v B$ in (23) and (24) and subsequently use the Maxwell equations with material sources (SI units):
\begin{eqnarray}
\nabla\cdot\v E=\,\frac{1}{\epsilon_0}(\rho-\nabla\cdot \v P),\qquad\qquad\;\nabla\cdot\v B=0, \\
\nabla\times \v E+\frac{\partial \v B}{\partial t}=0,\qquad\quad
\nabla\times \v B-\frac{1}{c^2}\frac{\partial \v E}{\partial t}= \mu_0\bigg(\v J+\nabla\times \v M+\frac{\partial \v P}{\partial t}\bigg),
\end{eqnarray}
where $\v P$ and $\v M$ are the polarisation and magnetisation vectors, to get the retarded fields \cite{12}:
\begin{eqnarray}
\v E =&\! - \nabla \bigg({1\over 4\pi\epsilon_0}\int\!\frac{[\rho-\nabla'\cdot\v P]}{R}\,d^3r' \bigg)
- \frac{\partial}{\partial t}\bigg (\frac{\mu_0}{4\pi}\int\frac{[\v J+\nabla'\times\v M+\partial\v P/\partial t]}{R}d^3r'\bigg),\nonumber\\
\\
\v B =& \nabla \times \bigg(\frac{\mu_0}{4\pi}\int\frac{[\v J +\nabla'\times\v M+\partial\v P/\partial t]}{R}\,d^3r'\bigg),
 \end{eqnarray}
or more compactly,
\begin{eqnarray}
\v E = - \nabla \Phi - \frac{\partial \v A}{\partial t},\quad\v B =\nabla \times \v A,
 \end{eqnarray}
 where we have defined the retarded scalar and vector potentials as
 \begin{eqnarray}
\Phi= {1\over 4\pi\epsilon_0}\!\int\!\frac{[\rho-\nabla'\cdot\v P]}{R}\,d^3r',\;\v A=\frac{\mu_0}{4\pi}\!\int\!\frac{[\v J +\nabla'\times\v M+\partial\v P/\partial t]}{R}\,d^3r'.
 \end{eqnarray}
Notice that if the vectors $\textbf{P}$ and $\textbf{M}$ are vanished in (33) and (34) then we recover (27) and (28).

In the last application  we write $\v F_1=c\v E$ and $\v F_2=\v B$ in (23) and (24), and subsequently use Maxwell's equations with magnetic monopoles expressed in Gaussian units as
\begin{eqnarray}
\nabla\cdot c\v E=\,4\pi c\rho_e,\qquad \qquad\;\nabla\cdot\v B=4\pi\rho_m, \\
-\nabla\times c\v E-\frac{\partial \v B}{\partial t}=4\pi\v J_m,\quad
\nabla\times \v B-\frac{1}{c}\frac{\partial \v E}{\partial t}= \frac{4\pi}{c}\v J_e,
\end{eqnarray}
where $\rho_e$ and $\v J_e$ are the electric charge and current densities and $\rho_m$ and $\v J_m$ are the magnetic charge and current densities,
to directly obtain the retarded fields with magnetic monopoles
\begin{eqnarray}
\v E = - \nabla\bigg( \int\!\frac{[\rho_e]}{R} \,d^3r'\bigg)
- \nabla \times \bigg(\int\frac{[\v J_m]}{R c}\,d^3r'\bigg)-\frac{1}{c} \frac{\partial}{\partial t}\bigg (\int\frac{\v [\v J_e]}{R c}\,d^3r'\bigg),\\
\v B = - \nabla \bigg(\int\!\frac{[\rho_m]}{R}\,d^3r'\bigg)
 +\nabla \times \bigg(\int\frac{[\v J_e]}{R c}\,d^3r'\bigg)- \frac{1}{c}\frac{\partial}{\partial t}\bigg (\int\frac{[\v J_m]}{R c}\,d^3r'\bigg),
 \end{eqnarray}
or more compactly:
\begin{eqnarray}
\v E = - \nabla \Phi_e  - \nabla\times \v A_m -\frac{1}{c} \frac{\partial \v A_e}{\partial t},\;\;\; \v B = - \nabla \Phi_m  + \nabla\times \v A_e - \frac{1}{c}\frac{\partial \v A_m}{\partial t} ,
\end{eqnarray}
where we have defined the retarded electric and magnetic scalar potentials as
 \begin{eqnarray}
\Phi_e=& \int\!\frac{[\rho_e]}{R}\,d^3r',\;\;
\Phi_m=\int\!\frac{[\rho_m}{R} \,d^3r',
 \end{eqnarray}
 and the retarded electric and magnetic vector potentials as
 \begin{eqnarray}
\v A_e=\int\frac{[\v J_e]}{R c}\,d^3r', \;\;\v A_m=\int\frac{[\v J_m]}{R c}\,d^3r'.
 \end{eqnarray}
Notice that if the magnetic densities $\rho_m$ and $\v J_m$ are vanished in (39) and (40) then we recover (27) and (28).

\section*{{{\fontfamily{qag}\selectfont {\large \textcolor[rgb]{0.00,0.00,0.49}{4. Pedagogical comment}}}}}
Standard textbook presentations of electromagnetism follow a different route to the electric and magnetic fields in the time-independent regime of Maxwell's equations than in the time-dependent regime of these equations. While in the time-independent regime, the Helmholtz theorem of the vector analysis is commonly applied to  find the electrostatic and magnetostatic fields in terms of their respective scalar and vector potentials, in the time-dependent regime Maxwell's sourceless equations are commonly used to introduce the scalar and vector potentials which are then inserted into Maxwell's  source equations, obtaining two coupled second-order equations involving potentials, which are shown to be  gauge invariant. The Lorenz-gauge condition is usually adopted to decouple these second-order equations and as a result, we arrive at two wave equations, which are solved to obtain the retarded scalar and vector potentials and by a subsequent differentiation of them we finally obtain the retarded electric and magnetic fields. It is evident that the usual method followed in the  time-dependent regime is somewhat more complicated than the usual method used in the time-independent regime which is based in the Helmholtz theorem. It is clear that for pedagogical reasons it is worth formulating an extension of the Helmholtz theorem, which may be useful in the time-dependent regime of Maxwell's equations. This has been made in the past and different extensions of the Helmholtz theorem to include the time-dependence of the fields have been formulated \cite{13,14,15,16,17,18,19,20,21,22,23,24,25,26,27,28}. But, as far as we are aware, none of these extensions of the Helmholtz theorem has yet been included in standard textbooks.

In our opinion, equations (23) and (24) for the retarded fields $\mathbf{F}_1$ and $\mathbf{F}_2$ could be included in a undergraduate presentation of Maxwell's equations. Similarly, the analogous equation but for a single retarded field \cite{18}:
\begin{eqnarray}
\v F= - \nabla \int\!\frac{[\nabla'\cdot\v F]}{4\pi R} d^3r'\! +\nabla\!\times \int\!\frac{[\!\nabla'\!\times\v F]}{4\pi R} d^3r'\! + \frac{1}{c^2}\frac{\partial}{\partial t}\int\!\frac{[\partial \v F/\partial t]}{4\pi R} d^3r',
 \end{eqnarray}
could alternatively be included in such an undergraduate presentation. We notice that if any of the fields $\mathbf{F}_1$ and $\mathbf{F}_2$ defined by  (23) and (24) is vanished then we obtain (44). The disadvantage of (23) and (24) is that they are somewhat complicated but their
advantage is that they \emph{directly} yield the retarded fields. In contrast, (44) has the advantage of being simpler but
the disadvantage is that it yields the retarded fields in an \emph{indirect} way.

\section*{{{\fontfamily{qag}\selectfont {\large \textcolor[rgb]{0.00,0.00,0.49}{5. Conclusion}}}}}

\noindent How to solve the time-independent Maxwell's equations? Answer: using the Helmhotz theorem of the vector analysis. How to solve the time-dependent Maxwell's equations? Answer: using the Helmholtz theorem for two retarded fields, which was formulated in this paper. The {\it key} to formulate this generalised theorem was the observation that the curls in the first regime are \emph{decoupled} quantities: $\nabla\times \v E=0$ and $\nabla\times \v B=\mu_0\v J$, while the curls in the second regime are \emph{coupled} quantities: $\nabla\times \v E+\partial \v B/\partial t=0$ and
 $\nabla\times \v B-(1/c^2)\partial \v E/\partial t=\mu_0\v J$. Accordingly, we formulated the Helmholtz theorem for two retarded vectors $\v F_1$ and $\v F_1$ in terms of their divergences: $\nabla\cdot\v F_\texttt{1}$ and $\nabla\cdot\v F_2$ and coupled curls: $-\nabla\times\v F_1-\partial \v F_2/\partial t$ and $\nabla\times\v F_2-(1/c^2)\partial \v F_1/\partial t$. The proof of the theorem required of the uniqueness of the solutions of the homogeneous wave equation, which was explicitly demonstrated in Appendix A. As applications, we applied the theorem to Maxwell's equations when they have electric charge and current densities, when they additionally have polarisation and magnetisation densities and when they additionally have magnetic charge and current densities. For each case we obtained the retarded fields in terms of their retarded potentials.  Standard pedagogy generally considers the Helmholtz theorem for a static field to {\it justify} the mathematical form of Maxwell's equations in the time-independent regime. Standard pedagogy might also consider the Helmholtz theorem for two retarded vector fields to {\it justify} the mathematical form of
 Maxwell's equations in the time-dependent regime.

 \appendix
\section*{{{\fontfamily{qag}\selectfont {\large \textcolor[rgb]{0.00,0.00,0.49}{Appendix A. Uniqueness of the solutions of the wave equation}}}}}

\noindent We have shown that the set of functions $\v F_1$ and $\v F_2$ given (11) satisfies (8) and (9). We have also pointed out that this set of functions is not generally unique and that the set formed by $\v F_1 +\v H_1$ and $\v F_2 +\v H_2$ also satisfies  Eqs.~(8) and (9) provided the functions $\v H_1$ and $\v H_2$ satisfy
\begin{eqnarray}
\qquad\;\;\;\;\;\nabla\cdot\v H_1=0, \qquad\qquad\;\; \,\,\nabla\cdot\v H_2=0,\\
-\nabla\times\v H_1-\frac{\partial \v H_2}{\partial t}=0, \quad \nabla\times\v H_2-\frac{1}{c^2}\frac{\partial \v H_1}{\partial t}=0.
\end{eqnarray}
The uniqueness of  $\v F_1$ and $\v F_2$ will be guaranteed if we are able to show that  $\v H_1$ and $\v H_2$ vanish everywhere. From (A.1) and (A.2) we obtain the homogeneous wave equations $\Box^2\v H_1=0$ and $\Box^2\v H_2=0.$ Our strategy is now to construct a relation that allows us to discover those initial and boundary conditions that guarantee that the only solutions of these wave  equations are the trivial ones:
$\v H_1=0$ and $\v H_2=0.$ We will see that these conditions are precisely those given in the formulation of the Helmholtz theorem for two retarded vector fields.

In order to proceed in a more rigorous way, we adopt the following notation: the Cartesian components of $\nabla\cF$ are  given by  $(\nabla\cF)^{ij}=\partial^i{\cal F}^{j}$. Latin indices $i,j,k...$ run from 1 to 3 and the summation convention $a_i a^i$  on repeated indices (one covariant and the other one contra-variant) is adopted. The normal derivative of $\cF$ at the surface $S$ (directed outwards from inside the volume $V$)
is denoted by the vector $\v n\!\cdot\!\nabla\cF$ and defined by its components as $(\v n\!\cdot\!\nabla\cF)^i=n_j\partial^j{\cal F}^i$ where $\v n$ is a unit vector outward to the surface  with $(\v n)^j=n^j.$  Now we can state the following uniqueness theorem for the solutions of the homogeneous vector wave equation:
\vskip 3pt
\noindent \textbf{Theorem.} If the vector function $\cF(\v r,t)$ satisfies the homogeneous wave equation $\Box^2\cF=0$,
the initial conditions that $\cF$ and $\partial\cF/\partial t$ vanish at $t=0$, and the boundary condition that
$\v n\!\cdot\!\nabla \cF$ is zero at the surface $S$ of the volume $V$, then $\cF(\v r,t)$ is identically zero.
\vskip 3pt
Proof. We write the wave equation $\Box^2\cF=0$ in index notation
\begin{eqnarray}
\partial_j\partial^j{\cal F}^i -\frac{1}{c^2}\frac{\partial^2 {\cal F}^i}{\partial t^2}=0.
\end{eqnarray}
We multiply this equation by $\partial {\cal F}_i/\partial t$ and obtain
\begin{eqnarray}
\frac{\partial {\cal F}_i}{\partial t}\bigg(\partial_j\partial^j{\cal F}^i -\frac{1}{c^2}\frac{\partial^2 {\cal F}^i}{\partial t^2}\bigg)=0.
\end{eqnarray}
The left-hand side can be re-written as
\begin{eqnarray}
\partial_j\bigg(\frac{\partial {\cal F}_i}{\partial t}\partial^j{\cal F}^i\bigg)-\frac{\partial}{\partial t}\bigg\{
\frac{\partial_j{\cal F}_{i}\partial^j{\cal F}^{i}}{2}+\frac{1}{2c^2}\frac{\partial{\cal F}_{i} }{\partial t}\frac{\partial{\cal F}^{i} }{\partial t}
\bigg\}=0.
\end{eqnarray}
The volume integration of this equation implies
\begin{eqnarray}
\frac{\partial}{\partial t}\int_V\bigg(\frac{\partial_j{\cal F}_{i}\partial^j{\cal F}^{i}}{2}+\frac{1}{2c^2}\frac{\partial{\cal F}_{i} }{\partial t}\frac{\partial{\cal F}^{i} }{\partial t}\bigg)\,d^3r=\int_V \partial_j\bigg(\frac{\partial {\cal F}_i}{\partial t}\partial^j{\cal F}^i\bigg)\, d^3r.
\end{eqnarray}
The volume integral on the right-side can be transformed into a surface integral,
\begin{eqnarray}
\frac{\partial}{\partial t}\int_V\bigg(\frac{\partial_j{\cal F}_{i}\partial^j{\cal F}^{i}}{2}+\frac{1}{2c^2}\frac{\partial{\cal F}_{i} }{\partial t}\frac{\partial{\cal F}^{i} }{\partial t}\bigg)\,d^3r=\oint_S\bigg(\frac{\partial {\cal F}_i}{\partial t}\bigg)\Big( n_j\partial^j{\cal F}^i\Big)\, dS.
\end{eqnarray}
From the boundary condition that $n_j\partial^j{\cal F}^i$ is zero at $S$ it follows that
 \begin{eqnarray}
\frac{\partial}{\partial t}\int_V\bigg(\frac{\partial_j{\cal F}_{i}\partial^j{\cal F}^{i}}{2}+\frac{1}{2c^2}\frac{\partial{\cal F}_{i} }{\partial t}\frac{\partial{\cal F}^{i} }{\partial t}\bigg)\,d^3r=0,
\end{eqnarray}
and therefore the integral is at most a function of space $g(x^i)$. The initial condition that
${\cal F}^{i}$ is zero at  $t=0$ implies the condition that $\partial^j{\cal F}^{i}$ is zero at $t=0$, which is used together with  the condition that $\partial{\cal F}^{i}/\partial t$ is zero at $t=0$ to show that $g(x^i)=0$. Accordingly,
\begin{eqnarray}
\int_V\bigg(\frac{\partial_j{\cal F}_i\partial^j{\cal F}^i}{2}+\frac{1}{2c^2}\frac{\partial_i{\cal F}_i }{\partial t}\frac{\partial{\cal F}^{i} }{\partial t}\bigg)\,d^3r =0,
\end{eqnarray}
or in the more familiar vector notation
  \begin{eqnarray}
\int_V\bigg(\frac{(\nabla\cF)^2}{2}+\frac{1}{c^2}\frac{(\partial\cF/\partial t)^2}{2}\bigg)\,d^3r =0.
\end{eqnarray}
Since the volume $V$ is arbitrary the integrand must vanish and considering that $\cF$ is a real function it follows that $\nabla\cF=0$ and $\partial\cF/\partial t=0$, which imply that $\cF=\;$ constant. Using the condition that $\cF$ is zero at $t=0$, it follows that this constant vanishes and then $\cF=0$, which proves the theorem.

The above theorem clearly describes a uniqueness theorem for the solutions of the homogeneous wave equation. Suppose we have two vector fields
$\cF_1$ and $\cF_2$ that satisfy the {\it same} wave equation $\Box^2\cF_1=0$ and  $\Box^2\cF_2=0$, the {\it same} initial conditions that $ \cF_1, \cF_2, \partial\cF_1/\partial t$ and $\partial\cF_2/\partial t$ vanish at $t=0$, and the {\it same} boundary conditions that
$\v n\!\cdot\!\nabla \cF_1$ and $\v n\!\cdot\!\nabla \cF_2$ are zero at the surface $S$ of the volume $V$. Now, let us write $\cF=\cF_2-\cF_1$. It follows that $\cF$ satisfies $\Box^2\cF=0$, the initial conditions that $\cF$ and $\partial\cF/\partial t$ vanish at $t=0$, and the boundary condition that $\v n\!\cdot\!\nabla \cF$ zero at the surface $S$. By the uniqueness theorem we have $\cF=0$ and therefore $\cF_1=\cF_2$, i.e., the solution is unique.

In particular, if the surface goes to infinity then we can assume the boundary conditions that $\nabla \cF$ and $\partial\cF/\partial t$
go to zero faster than $1/r^2$ as $r\rightarrow\infty$. Under these conditions, the surface integral in (A.7) vanishes. In this case we also arrive at (A.8) but with the volume $V$ extended over all space. Using the initial conditions that $\cF$ and $\partial\cF/\partial t$ vanish at $t=0$ we again imply (A.10) with the volume $V$ extended over all space and therefore  ${\cF}=0$ because $\nabla \cF$ and $\partial\cF/\partial t$ are zero in the limit
$r\rightarrow\infty$. But  ${\cF}=0$ means  $\cF_1=\cF_2$, i.e., the solution is unique.

We now return to the solutions $\v F_1 +\v H_1$ and $\v F_2 +\v H_2$ of (8) and (9). As previously noted: $\square^2\v H_1=0$ and $\square^2 \v H_2=0$. Therefore we can apply the uniqueness theorem to both $\v H_1$ and $\v H_2$ by assuming the initial conditions that $\v H_1, \v H_2, \partial\v H_1/\partial t, \partial\v H_2/\partial t$ vanish at $t=0$, the boundary conditions that
$\nabla \v H_1, \nabla \v H_2, \partial \textbf{H}_1/\partial t$ and $\partial \textbf{H}_2/\partial t$ go to zero faster than $1/r^2$ as $r\rightarrow\infty$, and the boundary conditions that $\v H_1$ and $\v H_2$ go to zero as $r\rightarrow\infty$. Under these conditions, the uniqueness theorem states that $\v H_1=0$ and $\v H_2=0.$

Therefore, the uniqueness of the vector functions $\v F_1$ and $\v F_2$ of the Helmholtz theorem for two retarded fields is guaranteed by assuming the initial conditions that $ \v F_1, \v F_2,
\partial\v F_1/\partial t$ and $\partial\v F_2/\partial t$  vanish at $t=0$, the boundary conditions that $\nabla \v F_1,\nabla \v F_2,
\partial\v F_1/\partial t$ and $\partial\v F_2/\partial t$ go to zero faster than $1/r^2$ as $r\rightarrow\infty$, and the boundary conditions that $\v F_1$ and $\v F_2$ go to zero as $r\rightarrow\infty$. These boundary conditions {\it imply} that the quantities $D_1, D_2, \v C_1$ and $\v C_2$, considered in the Helmholtz theorem for two retarded fields, go to zero faster than $1/r^2$ as $r\rightarrow\infty$.

\section*{{{\fontfamily{qag}\selectfont {\large \textcolor[rgb]{0.00,0.00,0.49}{References}}}}}

\end{document}